# Origin of frictional scaling law in circular twist layered interfaces: simulations and theory

Weidong Yan, Wengen Ouyang,[*] Ze Liu,[*]

*Department of Engineering Mechanics, School of Civil Engineering, Wuhan University, Wuhan, Hubei 430072, China*

[*]Corresponding authors. Email: w.g.ouyang@whu.edu.cn, ze.liu@whu.edu.cn

**Abstract:** Structural superlubricity based on twisted layered materials has stimulated great research interests. Recent MD simulations show that the circular twisted bilayer graphene (tBLG) presenting a size scaling of friction with strong Moiré-level oscillations. To reveal the physical origin of observed abnormal scaling, we proposed a theoretical formula and derived the analytic expression of frictional size scaling law of tBLG as $F \propto \theta^{-\frac{3}{2}} R^{\frac{1}{2}}$, where $\theta$ and $R$ are the interfacial twist angle and the radius of the flake, respectively. The predicted twist angle dependent scaling law agrees well with MD simulations and provides a rationalizing explanation for the scattered power scaling law measured in various experiments. Finally, we show clear evidence that the origin of the scaling law comes from the Moiré boundary, that is, the remaining part of the twisted layered interfaces after deleting the internal complete Moiré supercells. Our work provides new physical insights into the friction origin of layered materials and highlights the importance of accounting for Moiré boundary in the thermodynamic models of layered materials.

# 1 Introduction

When the contact between two atomic smooth layered materials rotated away from the commensurate stacking, the interlayer friction could be decreased to nearly zero (Dienwiebel et al., 2004; Hirano and Shinjo, 1990; Hirano et al., 1991; Peyrard and Aubry, 1983), a state known as structural superlubricity (SSL) (Hod et al., 2018; Liu et al., 2012). Because it provides a promising solution to the friction and wear problem in microelectromechanical system (MEMS), structural superlubricity has attracted extensive research interests in recent years (Cihan et al., 2016; Dienwiebel et al., 2004; Hod et al., 2018; Liu et al., 2012; Ouyang et al., 2018; Song et al., 2018; Yaniv and Koren, 2019). In particular, what is the source of friction and how is the interlayer friction dependence on the size in the superlubric state have become one of the focused research topics (Cihan et al., 2016; Dietzel et al., 2013; Dietzel et al., 2008; Dietzel et al., 2018; Koren and Duerig, 2016; Koren et al., 2015; Li et al., 2020; Liao et al., 2022; Mandelli et al., 2018; Qu et al., 2020; Varini et al., 2015; Wang et al., 2019a; Yaniv and Koren, 2019; Zhang et al., 2021). Experiments and simulations have shown that the contact edges make the main contribution to the interlayer friction (Liao et al., 2022; Mandelli et al., 2018; Qu et al., 2020; Wang et al., 2019a; Zhang et al., 2021), and the scaling law of friction with respect to the contact area is generally assumed as a power law ($F \propto A^\alpha$) for incommensurate contact between van der Waals (vdW) layered materials (Dietzel et al., 2013; Hartmuth et al., 2019; Koren and Duerig, 2016; Koren et al., 2015; Mandelli et al., 2018; Qu et al., 2020; Varini et al., 2015; Wang et al., 2019a). For example, it was found that the contact edge dominates the kinetic friction in superlubric systems and sublinear scaling laws ($\alpha < 1$) were observed due to the efficient cancellation of lateral forces at incommensurate interfaces. Recently, it was shown that the kinetic friction force of vdW layered heterostructures is dominated by a rim area consisting of incomplete Moiré tiles (MT) (Gigli et al., 2017; Koren and Duerig, 2016; Mandelli et al., 2017; Wang et al., 2019a), Wang et al. (Wang et al., 2019a) proposed an analytical formula for predicting the scaling law of kinetic friction for superlubric systems. It is noteworthy that both simulations and experiments show quite scattered $\alpha$ values, ranging from 0 to 0.5 (Cihan et al., 2016; Dietzel et al., 2013; Dietzel et al., 2018; Hartmuth et al., 2019; Koren et al., 2015; Özoğul et al., 2017; Qu et al., 2020). Theoretical studies attributed the scattering of $\alpha$ to the shapes of nanoclusters and its relative orientations to the 2D materials substrates (de Wijn, 2012; Dietzel et al., 2013; Varini et al., 2015). However, physical origin of the scattered scaling behavior remains unclear, a universal analytical approach for studying the scaling law of friction is highly desirable but still lacking.

In this work, we:

(i) Present an analytic expression of the interlayer potential (ILP) of twist graphene on an infinite graphene substrate as function of the twist angle and size (in the case of circular shaped graphene flake).

(ii) Develop a theoretical model to predict the scaling law of the interlayer friction between a circular twist graphene and an infinite graphene substrate as function of the twist angle and size, which shows scaling as $F \propto \theta^{-\frac{3}{2}} R^{\frac{1}{2}}$, which agrees well with MD simulation results and provides a rationalizing explanation for the scattered power scaling law measured in various experiments.

(iii) Show clear evidence that the origin of the scaling law comes from the Moiré boundary, that is, the remaining part of the twisted bilayer graphene after deleting the internal complete Moiré supercells.

(iv) Propose a new method to drastically tune the interlayer friction of layered vdW materials.

The outline of this paper is as follows. In section 2, using MD simulations, the interlayer potential energy of circular tBLG with different twist angle and radius are obtained, then based on spectrum analysis and starting from the lattice symmetry of graphene, we developed an analytical expression of the interlayer potential energy in rotational process. In section 3, we investigated the sliding energy barrier of circular graphene flakes sliding an infinite graphene substrate, and developed an analytical expression of the interlayer potential energy, which predicts a frictional scaling law as $F \propto \theta^{-\frac{3}{2}} R^{\frac{1}{2}}$, agreeing well with MD simulations. In section 4, based on systematical MD simulations, we revealed that the abnormal frictional scaling law originates from the Moiré boundary. Unlike the previous rough estimations ([Koren and Duerig, 2016](#); [Wang et al., 2019a](#); [Wang et al., 2019b](#)), we proposed a practical method to separate the inner and edge regions of vdW heterostructures at atomic precision, which allowed us to decouple their influence on the frictional properties of vdW heterostructures quantitatively. Applying this approach to the model system shows that the average energy barrier resulted from the edge atoms is about one order of magnitude larger than that of internal atoms, which is consistent to the recent experiments ([Qu et al., 2020](#)). In section 5, we considered a new method to drastically tune the interlayer friction of layered vdW materials.

## 2. Analysis the twist angle and size dependent interlayer potential energy of tBLG

### 2.1 MD simulation on the twist angle and size dependent interlayer potential energy and spectrum analysis

The MD simulation model consists of a rigid circular graphene flake rotating on an infinite graphene substrate which is rigid body as well, the infinite size of the graphene substrate is realized by applying periodic boundary conditions in lateral directions (Fig. 1a). The interaction between the flake and substrate are modeled as ILP ([Leven et al., 2014](#); [Leven et al., 2016](#); [Maaravi et al., 2017](#); [Ouyang et al., 2020a](#)) with a refined parametrization ([Ouyang et al., 2018](#)).

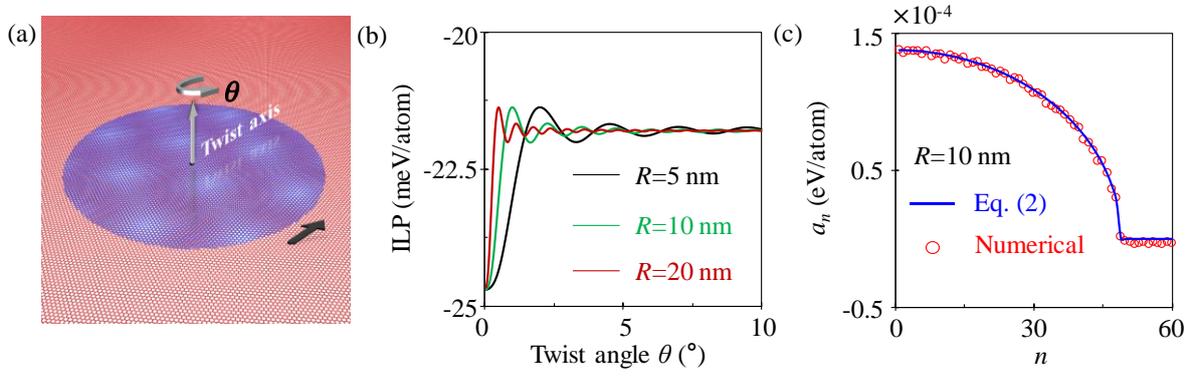

**Fig. 1** (a) MD simulation model of a circular graphene flake rotating on an infinite graphene substrate, where the periodic boundary condition is applied in laterally directions. (b) The simulated interlayer potential energy of typical graphene circular flakes with different radii as a function of the twist angle. (c) Fourier series coefficients of the interlayer potential energy of tBLG with a radius of 10 nm in (b).

We first investigated the in-plane rotation process at the equilibrium spacing (3.4 Å) in which the top graphene flake is rotated a twist angle of $\theta$ relative to the substrate with the rotation axis located at an AB- stacked position which is the center of the top graphene flake, and the initial configuration is AB-stacked. The simulated interlayer potential (ILP) as a function of twist angle $\theta$ is illustrated in Fig. 1b. It shows the ILP increases from the minimum value of $u_{min} \sim$ -24.3 meV/atom ($u_{min}$ corresponds to the averaged interlayer potential energy of AB-stacked bilayer graphene) to a constant -22.2 meV/atom with a strong oscillations, similar behavior has been observed before (Cao et al., 2022; Jiang et al., 2013; Morovati et al., 2022; Shibuta and Elliott, 2011; Zhang and Tadmor, 2017). We noted that the oscillations of the interlayer potential energy (e.g. the period) are closely related to the flake size (Fig. 1b), we therefore performed spectral analysis on the potential energy curves. Considering that both the graphene lattice and the Moiré superlattices show 6-fold rotation symmetry, we generally have

$$u(\theta) = \frac{a_0}{2} + \sum_{n=1}^{\infty} a_n \cos(6n\theta) \tag{1a}$$

where $a_0$ and $a_n$ are the coefficients of Fourier series. Interestingly, applying Eq. (1a) to the simulated interlayer potential energy reveals that $a_n$ rapidly approaches to zero as $n$ increases (Fig. 1c), which suggests that Eq. (1a) can be simplified as

$$u(\theta) = \frac{a_0}{2} + \sum_{n=1}^{n_{max}} a_n \cos(6n\theta) + o(n_{max}) \tag{1b}$$

where $n_{max}$ is the maximum number of summation terms. We found that the Fourier series coefficients $a_n$ can be well described by

$$\begin{cases} a_n = a_1 \sqrt{1 - \left(\frac{n}{n_{max}}\right)^2}, n \leq n_{max} \\ a_n = 0, n > n_{max} \end{cases} \tag{2}$$

Physically, considering the rotational symmetry of graphene lattice, the interlayer potential energy within $2\pi$ can be obtained by performing symmetry operation on the energy curve in the range of $[0, \pi/3]$, the maximum number $n_{max}$ essentially represents the highest fluctuation frequency of the interlayer potential energy within $[0, \pi/3]$, then we have $n_{max} = \lfloor \pi/(3\lambda) \rfloor$, where $\lambda$ is the period of the fluctuation, $\lfloor x \rfloor$ is floor function that gives as output the greatest integer less than or equal to $x$. Because the highest fluctuation frequency corresponds to the case that the displacement of any atom at the edge of the top graphene flake equals to half the lattice constant of graphene during rotation, that is, $\lambda R = \sqrt{3}a/2$, where $R$ is the radius of the rotational flake and $a \sim 0.246$ nm is the lattice constant of graphene. Hence, the maximum number $n_{max}$ reads

$$n_{max} = \left\lfloor \frac{2R\pi}{3\sqrt{3}a} \right\rfloor \tag{3}$$

Substituting Eq. (2) into Eq. (1b) we have

$$u(\theta) = \frac{a_0}{2} + \sum_{n=1}^{n_{max}} a_1 \sqrt{1 - \left(\frac{n}{n_{max}}\right)^2} \cos(6n\theta) \tag{4}$$

The summation term in Eq. (4) can be approximated by the corresponding definite integral within $n \in [1, n_{max}]$, then we have

$$u(\theta) = \frac{a_0}{2} + \int_0^1 a_1 n_{max} \cos(6n_{max}\theta x) \sqrt{1 - x^2} dx = \frac{a_0}{2} + \frac{a_1 \pi J_1(6n_{max}\theta)}{12\theta} \tag{5}$$

where, $J_1(\cdot)$ is the Bessel function of the first kind (Olver et al., 2010). Once the two parameters $a_0$ and $a_1$ are determined, the twist angle and size dependent interlayer potential energy can be readily obtained.

**2.2 Determination of the two parameters in the analytical formula of interlayer potential energy**

It is noteworthy that the simulated interlayer potential energy provides two boundary conditions (BCs), typical results are shown in Fig. 1b. The first one is that the averaged initial potential energy of tBLG with any size will equal to that of AB-stacked bilayer graphene. The other one is that if the radius of the top graphene flake tends to infinity, the interlayer potential energy converges to a constant value (Zhang and Tadmor, 2017) which is $u_0 \sim -22.2$ meV/atom (Fig. 1b). Letting $n_{max} \to \infty$, then Eq. (5) gives $a_0/2 = u_0$.

It should be pointed out that Fig. 1b is obtained by rotating the top graphene flake with its central axis passing through an AB-stacked position, to verify whether the two aforementioned BCs are still valid for different rotation axis, a series of simulations were performed (Fig. 2). Considering the translational symmetry of graphene lattice, only the triangular region in a carbon ring needed to be considered (the zoom-in of the red square frame in Fig. 2a). Seven rotation centers are chosen and numbered from 1 to 7 (Fig. 2a), the radius of the top graphene flakes is also varied ($R$ = 5, 10, and 20 nm). We observed that the ILP curves show similar behavior as Fig. 1b. Firstly, all the ILP curves converge to a constant value at large twist angle. Secondly, for any given rotation axis, the averaged interlayer potential energy shrinks to the same value at $\theta = 0$ and is independent on the size (Fig. 2c).

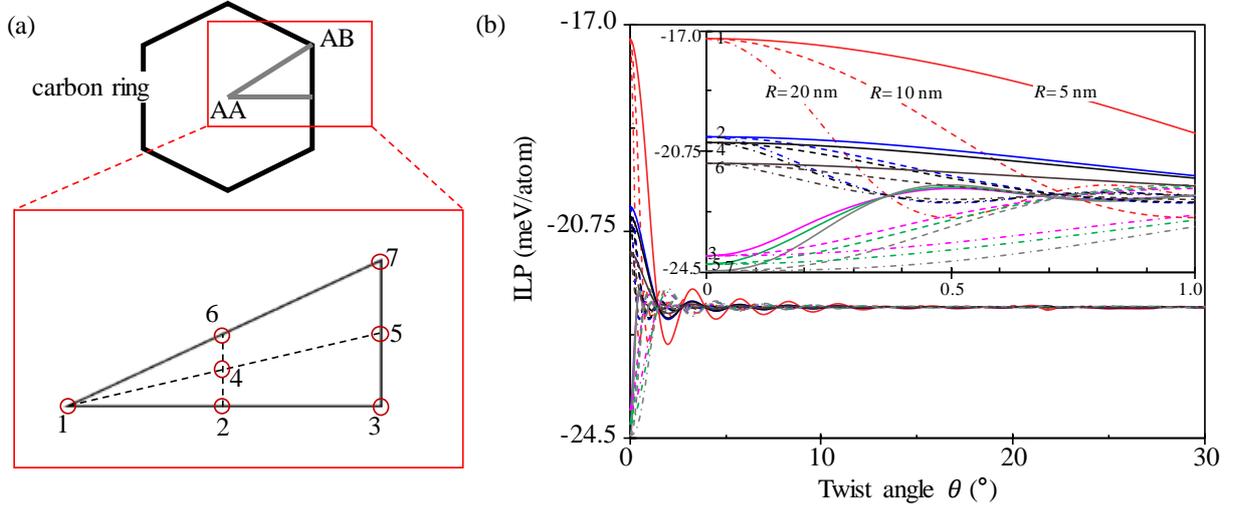

**Fig. 2** Dependence of the ILP curves on rotation centers and flake size. (a) Seven rotation centers are chosen from 1 to 7. (b) ILP curves for the seven rotation centers are obtained with MD simulations. The insert of (b) is Zoom-in the curves in the range of 0-1° in panel (b), where the numbers 1, 2, …,7 marked in the inset in panel (b) represent the seven rotation centers in (a).

Based on the first BC and Eq. (5), we have

$$u_{\theta=0} = \frac{a_0}{2} + \lim_{\theta \to 0} \frac{a_1 \pi J_1(6n_{max}\theta)}{12\theta} \tag{6}$$

where, $u_{\theta=0}$ is the averaged potential energy of bilayer graphene when twist angle is zero. Substituting Eq. (3) in Eq. (6) gives

$$a_1 = \frac{4}{\pi^2}\left(u_{\theta=0} - \frac{a_0}{2}\right)\left[\frac{2R}{3\sqrt{3}a}\right]^{-1} \tag{7}$$

## 2.3 Accuracy of the analytical formula of the interlayer potential energy for tBLG

As described in Secs. 2.1- 2.2, the ILP with different twist angle and size can be generally described by

$$u(\theta) = \frac{a_0}{2} + \frac{u_{\theta=0} - \frac{a_0}{2}}{n_{max}} \cdot \frac{J_1(6n_{max}\theta)}{3\theta} \tag{8}$$

To check the accuracy of the analytical formula (Eq. (8)), we carried out a series of MD simulations to compare with the analytical predictions, typical results are as shown in Fig. 3, where we show the ILP curves for the rotation center located at AB and AA-stacked positions, respectively. For the two initial configurations, $u_{\theta=0}$ equals to $u_{AA} \sim$ -17.3 and $u_{AB} \sim$ -24.3 meV/atom, respectively. It is clear that the analytical predictions agree well with the MD simulations (Fig. 3, the red diamonds and the blue lines correspond to the simulation results and the analytical predictions, respectively).

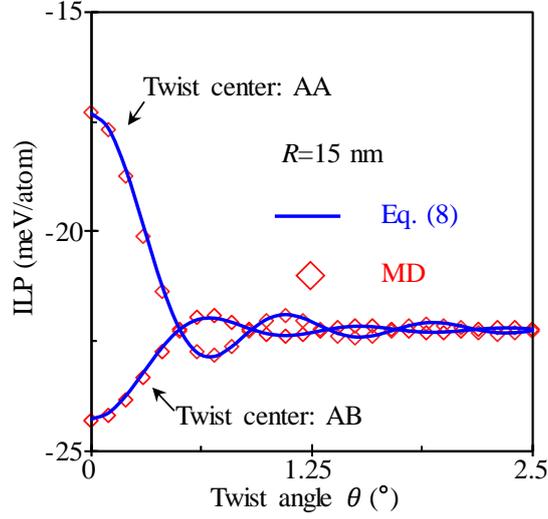

**Fig. 3** Comparison of ILP curves obtained by MD simulation and Eq. (8). The red diamonds are the results from MD simulations, and the blue lines are the results calculated by Eq. (8). The parameters used are $a_0/2$ = -22.2 meV/atom, $n_{max}$ = 73 corresponding to a circular flake with a radius of 15 nm. $u_{\theta=0}$ = -17.3 meV/atom and -24.3 meV/atom for the rotation center located at AA- and AB-stacked modes of tBLG.

## 3. Theoretical analysis the interlayer potential energy during translational motion

### 3.1 The analysis of the translational motion

In Sec. 2, we derived the analytical expression of the ILP of tBLG during interlayer rotation motion. In this section, we investigate the sliding potential energy surface (PES) of tBLG (averaged energy per atom) during translational motion. The previous study show that the sliding potential surface can be well approximated by (Gao and Huang, 2014; Lebedeva et al., 2010; Steiner et al., 2009; Verhoeven et al., 2004; Yan et al., 2022)

$$U(x, y, \theta) = U_1 + U_2 \cos(ky)\left[\cos(ky) + \cos(\sqrt{3}kx)\right] \quad (9)$$

where, $k = 2\pi/\sqrt{3}a$, $a$ is the lattice constant of graphene. $U_1$ and $U_2$ are two parameters to be determined, which depends on the twist angle.

Additionally, with two special twist centers when $(x, y) = (0,0)$ and $(0, a/\sqrt{3})$

$$\begin{cases} U(0,0,\theta) = U_1 + 2U_2 \\ U\left(0, \frac{a}{\sqrt{3}}, \theta\right) = U_1 - 0.25 U_2 \end{cases} \quad (10)$$

On the other hand, the twist angle dependent ILP for the two special twist centers (with initial AA- and AB-stacked configurations) are shown in Fig. 3, therefore we have (Eq. (8))

$$\begin{cases} U(0,0,\theta) = \frac{a_0}{2} + \frac{u_{AB} - \frac{a_0}{2}}{n_{max}} \cdot \frac{J_1(6n_{max}\theta)}{3\theta} \\ U\left(0, \frac{a}{\sqrt{3}}, \theta\right) = \frac{a_0}{2} + \frac{u_{AA} - \frac{a_0}{2}}{n_{max}} \cdot \frac{J_1(6n_{max}\theta)}{3\theta} \end{cases} \quad (11)$$

Substituting Eq. (11) into Eq. (10)

$$\begin{cases} U_1 = \frac{1}{9}\left[U(0,0,\theta) + 8U\left(0,\frac{a}{\sqrt{3}},\theta\right)\right] = \frac{a_0}{2} + \frac{(2u_{AB}+16u_{AA}-9a_0)}{54n_{max}\theta} \cdot J_1(6n_{max}\theta) \\ U_2 = \frac{4}{9}\left[U(0,0,\theta) - U\left(0,\frac{a}{\sqrt{3}},\theta\right)\right] = \frac{4(u_{AB}-u_{AA})}{27n_{max}\theta} \cdot J_1(6n_{max}\theta) \end{cases} \quad (12)$$

After $U_1$ and $U_2$ being determined, the sliding potential surface of tBLG with any twist angle and size can be analytically calculated (Eq. (9)). To show the accuracy of the analytical approximation, we take a tBLG with twist angle of 5° for example, it is clear that the theoretical calculation based on Eqs. (9) and (12) gives almost the same picture of the sliding potential surface as the MD simulation (Fig. 4). Remarkably, based on the analytical expression (Eq. (9)), the maximum energy barrier ($\Delta G_m$) during the translational sliding of the top graphene can be obtained, taking the sliding along the armchair direction of substrate as an example

$$\Delta G_m = \left| U(0,0,\theta) - U\left(0,\frac{a}{\sqrt{3}},\theta\right) \right| \quad (13)$$

where $U(0,0,\theta)$ and $U(0,a/\sqrt{3},\theta)$ correspond to the minimum and maximum values of ILP during sliding, respectively. Substituting Eq. (11) into Eq. (12) gives

$$\Delta G_m \approx \left| (u_{AA} - u_{AB}) \frac{\lambda}{\pi\theta} \cdot J_1\left(\frac{2\pi\theta}{\lambda}\right) \right| \quad (14)$$

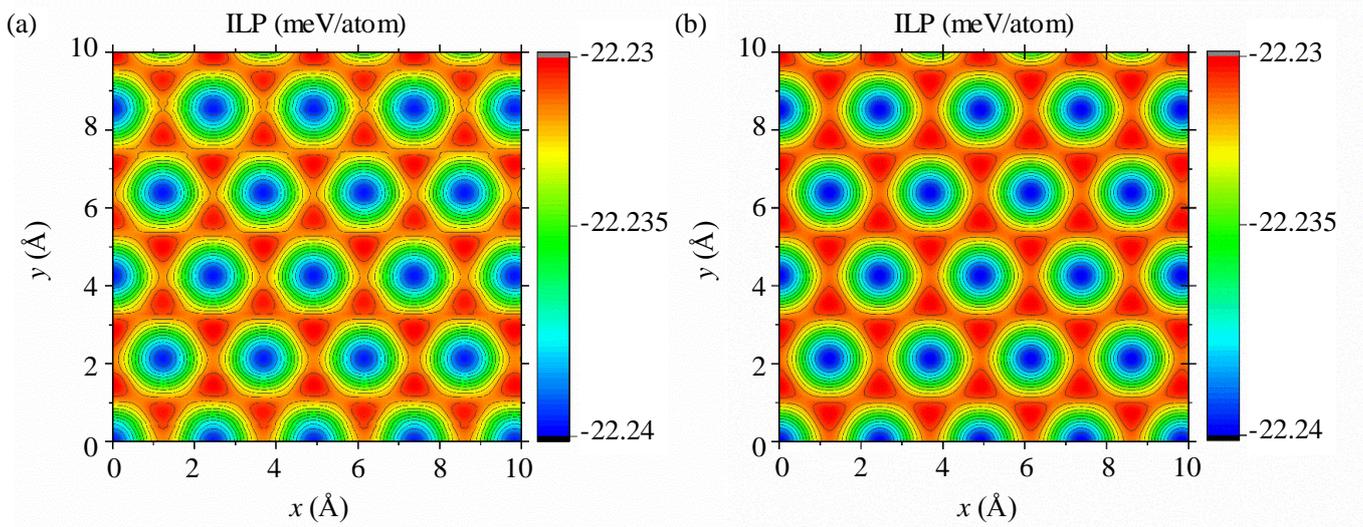

**Fig. 4** The typical sliding potential energy surface for tBLG with twist angle of 5°. The size of the top graphene flake is $R$ = 15 nm. (a) MD simulation results. (b) Theoretical results that calculated with Eq. (9), where $U_1$ = -22.2391 meV/atom and $U_2$ = -0.0017 meV/atom.

Equation (14) allows to investigate the twist angle and size dependent sliding energy barrier. For a given flake size, the maximum energy barrier drastically decreases as the bilayer graphene twists from the AB-stacked (Fig. 5a). Interestingly, there exists some twist angles where the maximum energy barrier is zero, in other words, the sliding PES is flat! On the other hand, if fixing the twist angle but changing the flake size, it is observed that the maximum sliding energy barrier grows with the size (or the contact area) (Fig. 5b), which presents a strong oscillation with a period of $a\sqrt{3\pi A}/(2\theta)$, where $a$ is the lattice constant of graphene, $A$ is the area of flake and $\theta$ is the twist angle. Remarkably, in both the cases, the predictions based on Eqs. (14) and (16) agree with the MD simulation results well (Fig. 5).

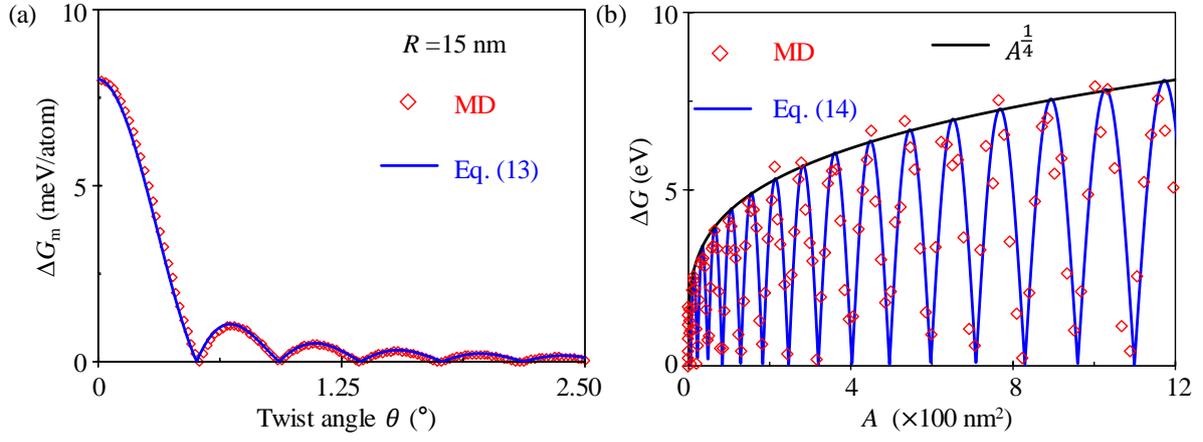

**Fig. 5** Theoretical prediction and MD simulation on the twist angle and size dependent sliding energy barrier. (a) The dependence of the maximum sliding energy barrier of a circular bilayer graphene (with a radius of 15 nm) on twist angle. (b) The contact area dependence of maximum sliding energy for circular tBLG with a fixed twist angle of 5°.

**3.2 Scaling law of interlayer friction**

Equation (14) is the averaged maximum sliding energy barrier, based on which the total sliding energy barrier of tBLG can be readily obtained as

$$\Delta G = \rho_s A \Delta G_m = \rho_s A \left| (u_{AA} - u_{AB}) \frac{\lambda}{\pi \theta} \cdot J_1\left(\frac{2\pi\theta}{\lambda}\right) \right| \tag{15}$$

where, $\rho_s$ is the atomic area density of graphene, and $A$ is the area of the flake. Applying the asymptotic cosine form of Bessel function at large values, we have

$$\Delta G = \rho_s A \Delta u_0 \frac{\lambda}{\pi \theta} \sqrt{\frac{\lambda}{\pi^2 \theta}} \left| \cos\left(\frac{2\pi\theta}{\lambda} - \frac{3}{4}\pi\right) \right| \tag{16}$$

where $\lambda R = \sqrt{3}a/2$ and $\Delta u_0 = u_{AA} - u_{AB}$. Since the maximum of $|\cos(\cdot)| = 1$, the maximum of Eq. (16) reads

$$\Delta G_{\max} = \alpha \theta^{-\frac{3}{2}} A^{\frac{1}{4}} \tag{17}$$

where $\alpha = \rho_s \Delta u_0 \frac{3a}{2\pi} \left(\frac{a^2}{12\pi}\right)^{\frac{1}{4}}$. The analytical expression (Eq. (17)) predicts that the envelope of the maximum sliding energy barrier is proportional to the area $A$ with a scaling exponent of 0.25, which agrees well with the previous MD simulation results (Koren and Duerig, 2016) and our own MD simulations results with a refined interlayer force field. Besides, Eq. (17) predicts that the envelope of the maximum sliding energy barrier is inversely proportional to $\theta^{\frac{3}{2}}$. Taking the tBLG with a fixed twist angle of 5° as example (Fig. 5b), Eq. (17) can perfectly predict the size dependence of the sliding energy barrier obtained by MD simulations, and it is noteworthy that there is no any fitting parameter in Eq. (17). Furthermore, the evolution of ILP and resistance force in the quasi-static sliding process are shown in Fig. 6a, where $R = 3.5$ nm and $\theta = 5°$. When fixed the twist angle but varied the size, it is observed that the static friction force is directly proportional to the maximum sliding energy barrier in the sliding process (Fig. 6b). We therefore conclude that the scaling law in

Eq. (17) also applies to the scaling of interlayer friction of tBLG.

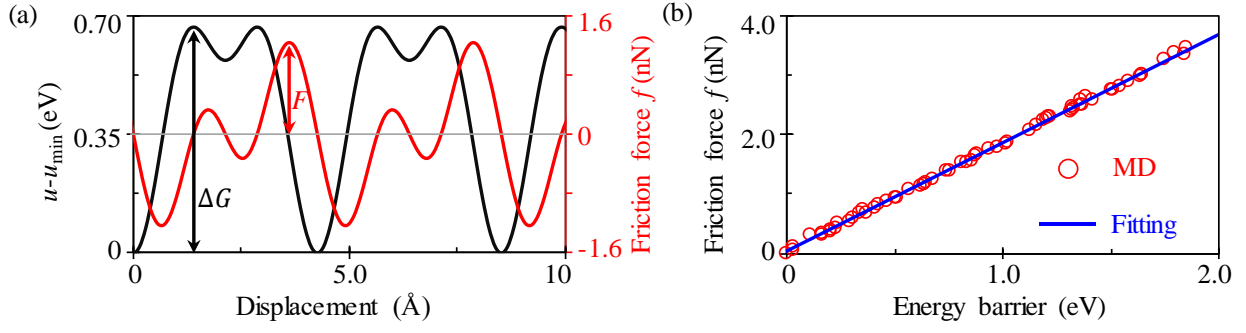

**Fig. 6** (a) The evolution of ILP and resistance force during the quasi-static sliding process. The maximum energy barrier and the static friction force are denoted as $\Delta G$ and $F$, respectively. In the calculation, $R = 3.5$ nm and $\theta = 5°$. (b) The relationship between the static interlayer friction and the sliding energy barrier, where circular graphene flakes have the same twist angle of 5° but different size $R$ (from 0 to 10 nm). The sliding direction is along the armchair direction of the substrate.

## 4. The origin of the friction scaling law

The above analysis clearly shows that the maximum sliding energy barrier or the static interlayer friction of circular tBLG scales with the area as $F \propto A^{1/4}$ (Eq. (17)), which contradicts the classic friction law, where it is either proportional to the contact area ($\propto A^1$) or to the size of the edge ($\propto A^{1/2}$). For instance, in previous study, it has been revealed that the edge atoms can significantly leads to the energy barrier or non-zero friction of tBLG since the dangling bond pinning effect of the edge atoms, or the edge atoms are more prone to fluctuate in the sliding process due to the weaker constraints compared with the atoms located in internal region (Koren and Duerig, 2016; Qu et al., 2020; Wang et al., 2019a). In this section, we will investigate the physical origination of the abnormal scaling law (Eq. (17)).

### 4.1 The Moiré boundary

Considering that Moiré superlattices generally form at twisted interfaces of van der Waals materials, which plays an important role on their electrical (Cao et al., 2018) (Koren et al., 2016), thermal (Ouyang et al., 2020b), mechanical (Ouyang et al., 2021) and tribological properties (Song et al., 2018), the abnormal scaling law could be from the Moiré effect. Actually, recent studies have shown that the incomplete Moiré superlattices near the flake edge, i.e., MTs, can dominate the frictional scaling behavior of layered materials (Koren and Duerig, 2016; Wang et al., 2019b). To decouple the friction contribution of the incomplete MTs, for a given tBLG, the internal region with all complete MTs should be distinguished first. As is shown in Fig. 7a-b, each diamond (surrounded by the blue dashed lines) represents the minimum unit of a Moiré superlattice in tBLG with twist angle of 1°. The diamonds are defined as MTs, and each cross point of a MT is defined as its vertex, such as $P_1$ in Fig. 7a. We used these vertexes to distinguish (in)complete MT quantitatively, that is, only when all four vertexes of a MT locate in a region enclosed by the geometric edge (the red solid line in Fig. 7a-b), the MT is considered to be a complete one and be counted as part of the internal region, typically such as the red diamond $L_1$ shown in Fig. 7a. Otherwise the MT is counted as part of the edge region. To distinguish with the geometric edge, we define the boundary of internal region as Moiré boundary (MB, the green solid line). Unlike the geometric edge, the shape of MB depends on the relative position (Fig. 7a-b) and orientation (Fig. 8a-b) of the top graphene flake with respect to the graphene substrate, in other words, MB is intrinsically a dynamic edge. For example, when the top graphene moves 1.5 Å along the $x$ direction, one of vertexes (i.e. $P_1$ of MT $L_1$) will move outside of its geometric edge, then the MT $L_1$ is not part of the internal region anymore (Fig. 7b). Meanwhile, the '$W$' shaped internal region changes to the '$Z$' shape (Fig. 7a-b). When outputting the interlayer potential energy of the internal and the edge regions (Fig. 7c-d, the red and green dots are the ILP of the internal and edge regions, respectively), it is observed that the energy barrier in the sliding process is almost completely contributed from the edge region. The energy barrier of the edge region is $\Delta G_e \sim 0.09$ meV/atom, far larger than that of the internal region ($\Delta G_i \sim 0.002$ meV/atom).

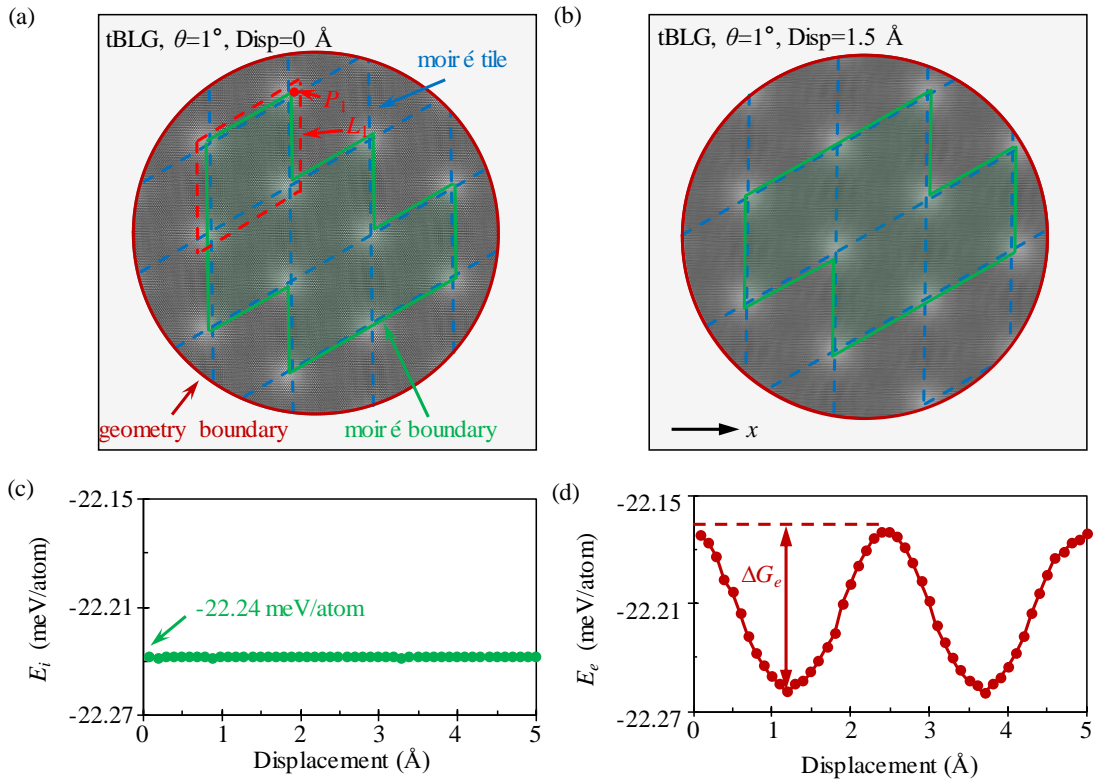

**Fig. 7** The evolution of Moiré superlattices and interlayer potential energy during translational motion. (a) The initial configuration of a tBLG with radius of $R = 25$ nm and twist angle of 1°. (b) The configuration after the top circular graphene flake sliding along the $x$ direction with a displacement of 1.5 Å. The red solid lines represent the geometry edge of the top graphene. The grids formed with blue dotted lines are Moiré unit-cells. The green solid lines mark the Moiré boundary and the green area is defined as the internal region. The region enclosed by the Moiré boundary and geometry edge is defined the edge region. The red dotted line in (a) represents a complete Moiré tile $L_1$ which belongs to the internal region and $P_1$ is a vertex of the Moiré tile. (c)-(d) Evolution of interlayer potential energy contributed from the internal region (c) and edge region (d) of the tBLG during sliding process along the $x$ direction.

Similarly, the evolution of Moiré superlattices and Moiré boundary during rotational motion can also be quantified (Fig. 8), the increasing of twist angle leads to the decreasing of Moiré pattern size (the grids enclosed by the blue dashed lines in Fig. 8a-b). Fig. 8c-d show the ILP contributed from the internal and edge regions, respectively. It is again observed that the fluctuation of ILP in the internal region is negligible by comparison with that of ILP in the edge region.

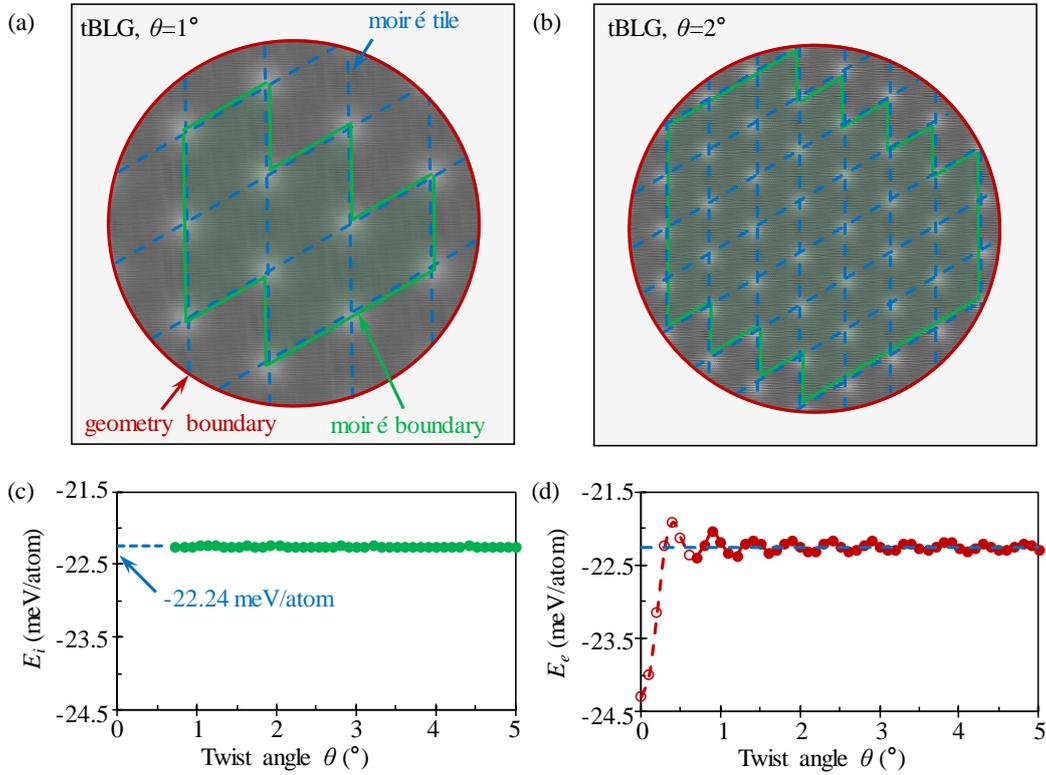

**Fig. 8** The evolution of Moiré superlattices and interlayer potential energy during rotational motion. (a) The initial configuration of tBLG with radius of $R = 25$ nm and twist angle of 1°. (b) The configuration after the top circular graphene rotating 1° further. The red solid lines represent the geometry edge of the top graphene. (c)-(d) Evolution of Interlayer potential energy contributed from the internal region (c) and edge region (d) of the tBLG during the rotation motion. The dashed line in (c) and the empty circles in (d) represent the twist angle range of smaller than 0.7°, where the flake (with the radius of 25 nm) is too small to include a complete Moiré tile. Then, the whole flake belongs to the edge region, accordingly the rotation energy barrier in this range origins from the edge region.

## 4.2 Moiré boundary versus Geometric edge

The geometric edge is widely used to distinguish edge atoms and study the edge effect (Liao et al., 2022; Liu et al., 2012; Mandelli et al., 2018; Qu et al., 2020; Wang et al., 2019a; Zhang et al., 2013; Zheng et al., 2008). Considering that the geometric edge is not continuous and smooth, geometric edge with very small width is usually used. To compare the rationality of the Moiré boundary and the geometric edge in the studying interlayer friction, we defined geometry edge as a rim region with a fixed width $t \sim 1$ Å (Fig. 9a-b). It is found that by distinguish the internal region with MB, the energy barrier of the internal region depends weakly on the twist angle (the green lines) and it keeps one order of magnitude lower than that of the edge region (the red lines) for both circular flake (Fig. 9c) and square flake (Fig. 9d). However, if using the geometry edge to distinguish the internal region, the contribution to the energy barrier from the internal and edge regions cannot be decoupled (Fig. 9e-f), for instance, the interlayer potential energy from both the internal and edge regions show a strong shape dependence (Fig. 9c-d), i.e., 60°- and 60°-symmetry for circular and square flake, respectively. In contrast, the Moiré edge approach always show a 60°-symmetry dependence of friction on the twist angle, which agrees well with the experimental observations (Dienwiebel et al., 2004; Liu et al., 2012; Song et al., 2018). Therefore, we conclude that the definition of MB is a more suitable for decoupling the edge effect in studying the frictional behavior of twisted bilayer materials.

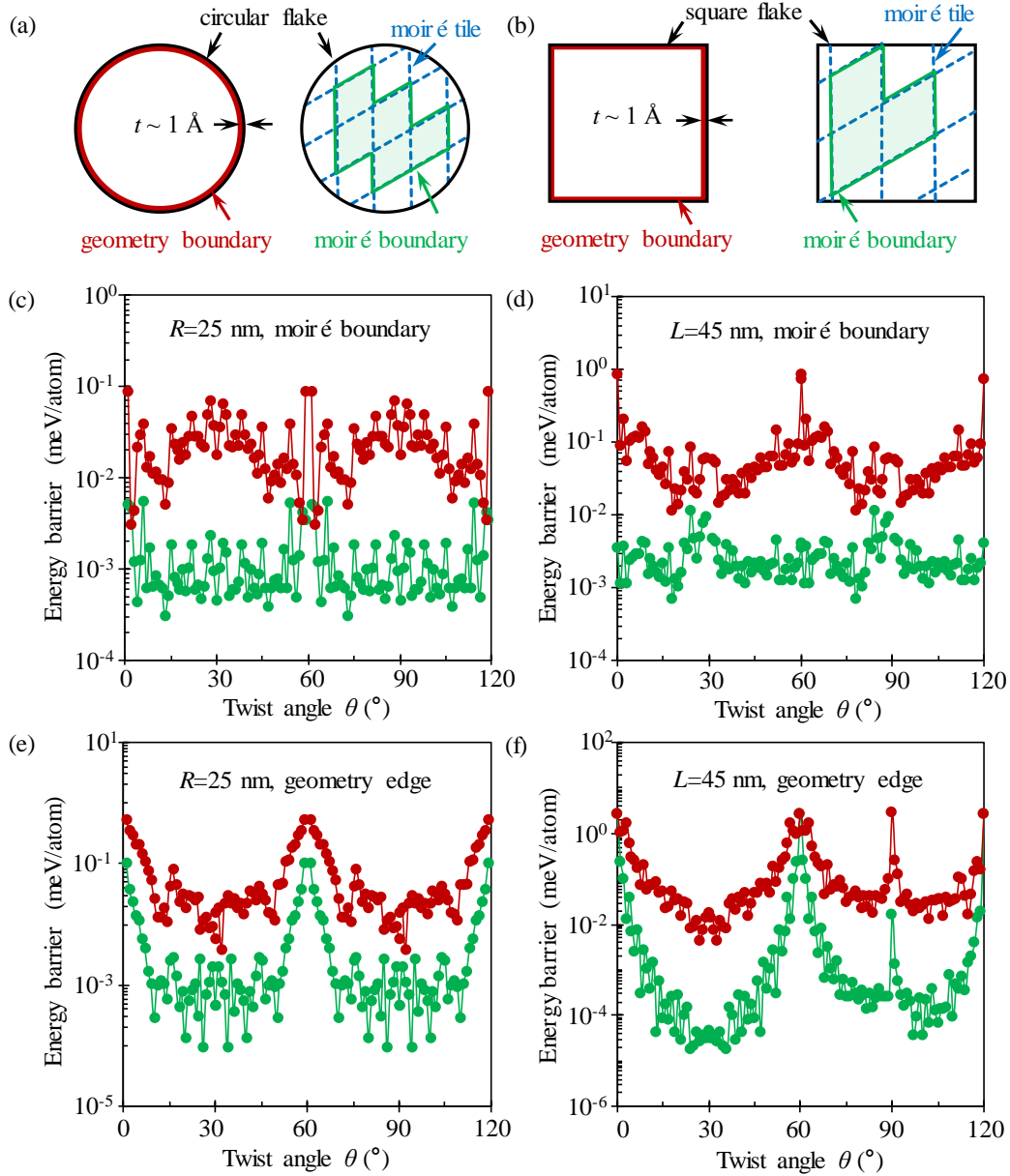

**Fig. 9** Edge effect in the calculation of twist angle dependence of interlayer potential energy. (a)-(b) Sketches the Moiré edge and the geometry edge. An internal region can be distinguished either by the Moiré boundary (c)-(d) or by the geometry edge (e)-(f) for both circular flake (c, e) and square flake (d, f). In the MD simulations, the radius of the circular flake is 25 nm, and the side length of the square flake is 45 nm. To find the maximum energy barrier quickly, the sliding direction for all twist angles is along armchair direction of substrate.

The above insights into the origination of energy barrier indicates that the discovered abnormal scaling law (Eq. (17)) is completely from the edge region as identified by the MB. To demonstrate this quantitatively, we calculated the size scaling of the energy barrier for circular tBLG with twist angle of 5° (Fig. 10) using both the theoretical model (blue solid line) and MD simulations (black triangles). We further extract the contribution of energy barrier from the internal (green circles) and edge region (red squares) that separated by the MB in MD simulations, it is found that the energy barrier is almost completely contributed from the Moiré edge, and all the calculated maximum energy barriers exactly locate at the predicted grey curve (Fig. 10).

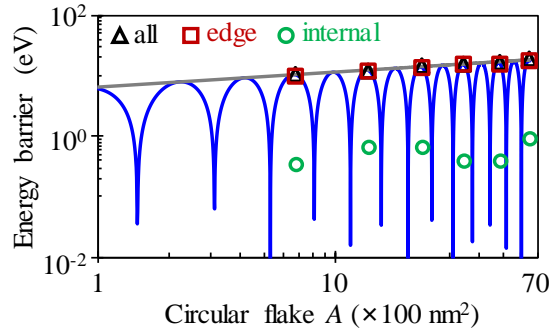

Fig. 10 The origin of the discovered abnormal scaling law. The blue line is obtained from Eq. (16) and the grey line is the envelope of energy barrier predicted by Eq. (17). Six different sizes which are located at the envelope are selected to calculate the contribution of the internal and edge regions to the energy barrier of tBLGs. The black triangles are the total energy barrier. Red squares and green circles are the energy barriers contributed from the edge region and the internal region, respectively.

## 5. Tuning interlayer friction by applying the MB

We show above that the edge region identified by the MB dominates the interlayer friction of twist bilayer materials, it is straightforward that if tailoring the shape of tBLG according to the MB, the interlayer friction can be drastically reduced. Typical results are shown in Fig. 11, if designing the shape of the top graphene flake according to the Moiré superlattices, such as the '$W$' shape (inset of Fig. 11), by comparison with the initial circular shape, the sliding energy barrier of the ILP is significantly smaller (the red and black lines in Fig. 11). More specifically, the sliding energy barrier of the circular shape graphene and the '$W$' shape graphene are 0.05 meV/atom and 0.007 meV/atom, respectively, i.e., the designed configuration reduces the static interface friction by about one order of magnitude.

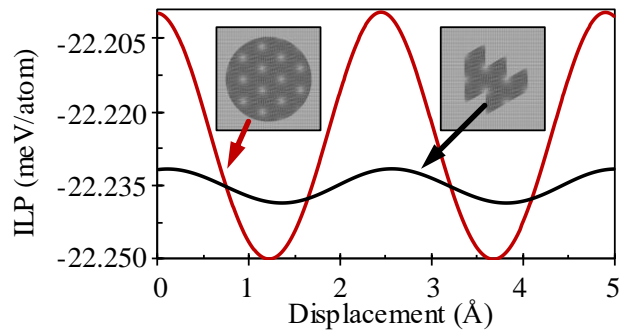

Fig. 11 Tuning interlayer friction by designing the shape of tBLG. Calculated interlayer potential energy of a circular shape tBLG and a 'W' shape tBLG (insets) with the same twist angle ($\theta = 5°$). The red and black lines correspond to the ILP of the circular and 'W' shaped tBLG, respectively. In the MD simulations, the sliding direction is fixed along armchair direction of substrate.

**Conclusions**

In summary, we studied the model system of a rigid circular graphene flake sliding on an infinite graphene substrate using the state-of-art force field and show that the variation of the interlayer potential energy with respect to the twist angle can be described using a Bessel function. Based on this finding, we derived an analytical formula which predicts an abnormal friction scaling law, that is, the static friction ($F$) scales with the contact area ($A$) as $F \propto A^{1/4}$. By defining a new Moiré boundary to distinguish edge atoms, the frictional scaling law is found to be dominated by the dynamic Moiré edge. Moreover, we proposed a method to tuning the interlayer friction of twist bilayer materials by tailoring the shape according to the Moiré boundary, based on which the interlayer friction can be reduced by one order of magnitude. Our work provides not only a theoretical framework to quantify the dependence of energy barrier (and friction) on the twist angle and size, but also new insights into the understanding and tuning of interlayer friction of layered vdW materials, which points out a possible way for realizing larger-scale superlubricity.